\documentstyle[aps,psfig,floats,twocolumn,prb]{revtex}
\begin{document}

\renewcommand{\footnoterule}{}
\textfloatsep=0.5cm
\intextsep=0.5cm

%
\newcommand{\siloo}{Si(001)-2$\times$1\ }
\newcommand{\hm}{H$_2$}
\newcommand{\si}[2]{Si$_{#1}$H$_{#2}$}
%
%
\newcommand{\jour}[4]{#1\ {\bf #2},\ #3\ (#4)}
\newcommand{\Rref}[1]{Ref.~\onlinecite{#1}}
\newcommand{\Fref}[1]{Fig.~\ref{#1}}
\newcommand{\Tref}[1]{Table~\ref{#1}}
\newcommand{\etal}{{\em et al. }}
\newcommand{\PRB}{Phys. Rev. B}
\newcommand{\PRL}{Phys. Rev. Lett.}
\newcommand{\JCP}{J. Chem. Phys.}
\newcommand{\SS}{Surf. Sci.}
%

\ifpreprintsty
\addtolength{\textwidth}{-1.5cm}
\addtolength{\oddsidemargin}{-1.0cm}
\addtolength{\evensidemargin}{-1.0cm}
\else
\twocolumn[\hsize\textwidth%
\columnwidth\hsize\csname@twocolumnfalse\endcsname
\fi
\tightenlines
\draft

\title{Effect of the cluster size in modeling the H$_2$ desorption and
  dissociative adsorption on Si(001)}

\author{E. Penev, P. Kratzer and M. Scheffler}
\address{Fritz-Haber-Institut der Max-Planck-Gesellschaft, Faradayweg 4-6, 14195 Berlin-Dahlem, Germany}

\date{\today}

\maketitle

\begin{abstract}
Three different clusters, \si{9}{12}, \si{15}{16}, and
\si{21}{20}, are used in density-functional theory  calculations 
in conjunction with {\it ab initio} pseudopotentials to study how the energetics of \hm\ dissociative adsorption on and associative desorption from Si(001) depends on the cluster size. 
The results are compared to five-layer slab calculations using the same pseudopotentials and high quality plane-wave basis set. Several exchange-correlation functionals are employed. 
Our analysis suggests that the smaller clusters generally overestimate the
activation barriers and reaction energy. The \si{21}{20} cluster, however, is
found to predict reaction energetics, with $E_{a}^{\rm des}=56 \pm 3\mbox{
  kcal/mol }(2.4\pm 0.1\mbox{ eV})$, reasonably close (though still different) to that obtained
from the slab calculations. Differences in the calculated
activation energies are discussed in relation to the efficiency of 
clusters to describe the properties of the clean \siloo surface.
\end{abstract}

\pacs{PACS numbers: 68.35.-p, 82.65.My}

\ifpreprintsty\else\vskip1pc]\fi
\narrowtext

\vskip2pc

\section{Introduction}

Cluster models are a frequently used tool for
studying different aspects of physics and chemistry of clean surfaces, 
adsorption, and surface chemical reactions. 
An example which attracted 
considerable experimental and theoretical interest over the last years is the
H dissociative adsorption and associative desorption on the \siloo
surface~\cite{Doren96} which is a subject of this study. The intriguing
experimental result that the \hm\ desorption from Si(001) follows first-order
kinetics~\cite{Sinniah89,Wise91,Hofer92} has triggered an intense theoretical
activity in this field mainly concentrated on the mechanism(s) leading to such
an unusual behavior. The available first principles calculations address the
latter question on the basis of two different models: ({\it i}) the cluster
approximation using either configuration-interaction (CI) methods~\cite{Nachtigall93,Nachtigall91,Wu93,Nachtigall94,Jing93,Jing95,Nachtigall96,Radeke96}
or density-functional theory (DFT)~\cite{Nachtigall94,Pai95} to
describe the exchange and correlation effects or ({\it ii}) extended slab models for the \siloo surface using DFT~\cite{Vittadini94,Pehlke95,Kratzer95}. 

The DFT slab calculations all agree in their conclusions supporting the {\it
  pre-pairing} mechanism~\cite{Wise91} according to which two hydrogens are
pre-paired on the same Si surface dimer and associatively desorb through an
asymmetric transition state (TS). The cluster calculations, however, have led
to different conclusions.  All these calculations find a rather high barrier for desorption of two hydrogens from a single Si dimer,
e.g. $E_{a}^{\rm des} =$ 74--75 kcal/mol (3.2 eV)~\cite{Nachtigall94},
85--86 kcal/mol (3.7 eV)~\cite{Jing95}, 82--85 kcal/mol (3.6--3.7 eV)~\cite{Radeke96}. Comparing to the experimental activation energy 
of desorption of $\sim 58$~kcal/mol (2.5 eV)~\cite{Wise91,Hofer92,Flowers93,Bratu96}, these findings were interpreted as being compelling evidence against the pre-paring mechanism.
In an attempt to reconcile the experimentally observed energetics and kinetics
of desorption from the monohydride phase, various defect-mediated mechanisms~\cite{Nachtigall91,Wu93,Nachtigall94,Radeke96} were
suggested including formation of metastable dihydride species as an intermediate step. 
All the above studies have in common that their argumentation rests on
computational schemes based on CI and the small and simple \si{9}{12} cluster
to model the \siloo  surface except the work by Nachtigall~\etal\cite{Nachtigall94} where DFT has been employed. In some studies larger clusters, like
\si{16}{x}~\cite{Nachtigall94} or \si{21}{x}~\cite{N&D}, have been
used. Nevertheless, the effect of cluster size on the energetics of \hm\
adsorption/desorption on Si(001) has not been analyzed in detail in the literature. 

The only cluster-based support to direct desorption via the pre-pairing
mechanism comes from the DFT calculations by Pai and Doren~\cite{Pai95}.
Using the non-local Becke-Lee-Yang-Parr (BLYP) functional~\cite{bX,lypC}, they
find $E_{a}^{\rm des} = 64.9\mbox{ kcal/mol}$ (2.8~eV) including a zero-point
energy (ZPE) correction. Given the uncertainties originating both from the use
of different functionals and from inherent limitations of the  cluster approximation,
they considered their result to be compatible with the experimental desorption
energy. The first possible source of error, the reliability of the functional,
was addressed by Nachtigall~\etal\cite{Nachtigall96}, who systematically
compared various density functionals to the results of the state-of-the-art
computational tool in quantum chemistry, an extrapolated quadratic CI method.
For the sake of computational feasibility, they concentrated on a few simple
test cases, four reactions involving silanes, and a \si{2}{6} cluster with a
geometry chosen to mimic \hm\ desorption from Si(001). While the extrapolated
quadratic CI method gives a reference value of 90.4~kcal/mol (3.92~eV) for $E_{a}^{\rm des}$, the Perdew-Wang (PW91) functional~\cite{pw91} underestimates
$E_{a}^{\rm des}$ by 9.5~kcal/mol (0.41~eV) compared to this reference. The
Becke-Perdew (BP) functional~\cite{bX,pC} gives an even lower barrier,
$E_{a}^{\rm des} = 79.4\mbox{ kcal/mol}$ (3.44~eV). Generally, Nachtigall~\etal find the B3LYP
functional~\cite{b3lyp} to give closest agreement with their CI calculations,
while the BLYP functional is performing second best. A similar trend
concerning the performance of different functionals was also observed in the
case of H diffusion on the \siloo surface~\cite{Nachtigall95}. However, we
find it not to be {\it a priori} clear whether their conclusions could be
applied to better cluster approximations to the Si(001) surface, because the
electronic wave functions at a surface are generally more extended than in a
cluster and this may naturally affect the electron-electron correlations. It
is one of the aims of the present paper to study the performance of different functionals with the size of the clusters. 

A second issue which hampers a clear-cut comparison of different approaches
lies in the geometric structure of the clean Si(001) surface.  
While the DFT calculations give the correct description of the geometry 
of the clean surface, normally a $p(2\times 2)$ or $c(4\times 2)$
reconstruction with buckled Si dimers~\cite{Dabrowski92,Wolkow92}, 
CI calculations predict a symmetric ground state of the clusters. 
This could imply substantial differences between the two approaches in the
surface relaxation during the adsorption/desorption process as well as in the
surface electronic structure. Since \hm\ dissociation on Si(001) is known to
couple to the surface motion~\cite{Brenig,Pehlke95,Kratzer95}, one could
expect these differences in the description of the clean Si(001) surface also
to influence the reaction energetics predicted by different calculational approaches.

In the light of the above we find of particular interest to bridge the DFT
slab and cluster calculations by studying the convergence of the
\hm\ adsorption/desorption energetics with cluster size. 
Since the results of the DFT calculations agree about a direct 
desorption process and the very recent experiment by
Flowers~\etal\cite{Flowers98} also supports this mechanism, we shall perform our analysis adopting the pre-pairing scenario for the
Si(001) surface. A description of the computational method used in the present
work is given in the next section. In Sec.~\ref{sec:III} particular attention
is paid to the structure of the clean Si(001) surface. Clusters with one, two
and three surface dimers are used in Sec.~\ref{sec:IV} to study the energetics
of the adsorption/desorption process. A summary of the results and discussion
is presented in the last section.


\section{Calculations}
\label{sec:II}

\subsection{The systems used for modeling \hm/Si(001)}

Generally, cluster models for surface chemical processes are treated in
conjunction with a basis set for the electronic states consisting of localized
orbitals. However, special care is required in choosing a basis set which
meets the desired level of accuracy. In order to isolate the different
approximations inherent in the choice of the cluster and the basis set
size, we decide to perform total energy calculations within the
DFT scheme as implemented in the {\tt fhimd} package~\cite{Bockstedte97},
employing a plane waves basis set with well-controlled convergence properties.
We report results for the three different clusters shown in~\Fref{fig:1}~(a),
\si{9}{12}, \si{15}{16} and  \si{21}{20}. Each of them represents the topmost four layers of the reconstructed surface. They differ by 
the number of Si-Si surface dimers they contain. The larger \si{15}{16}
and \si{21}{20} clusters are derived from \si{9}{12} by adding one and
two dimers, respectively, in the $[1{\overline 1}0]$ direction 
along with their full coordination of second layer Si atoms, one third and one
fourth layer Si atom. All dangling bonds of the silicons in the subsurface
layers are saturated with hydrogen atoms.
By increasing the cluster size in this way we aim at reaching three ascendingly
improved approximations to the clean Si(001) surface: \si{9}{12} is the minimal
one to represent the symmetric $2\times 1$ reconstruction; \si{15}{16} is
the smallest cluster that enables to model the $p(2\times 2)$ surface 
reconstruction. Finally, \si{21}{20} contains one surface dimer surrounded by
two others, thus having the same local environment as on the surface. 
In the context of \hm\ desorption from the monohydride phase, 
the larger clusters also allow to study the interaction between adjacent 
occupied dimers.

%
\begin{center}
\begin{figure}[h]
\ \psfig{figure=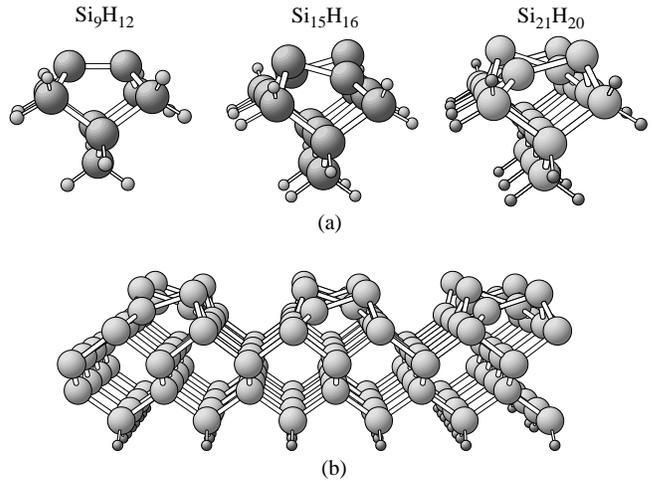,width=8.5cm}
\vskip0.5cm
\caption{ BP optimized geometries of (a) the clusters used to model the \siloo
  surface and (b) the five-layer $p(2\times2)$ slab used as a reference system.
\label{fig:1}}
\end{figure}
\end{center}

Since the super cell approach implemented in {\tt fhimd} implies periodic 
boundary conditions for the wave functions, the clusters are placed in a
sufficiently large orthorhombic unit cell, where they are separated by
$\approx 7$ \AA\ to avoid unwanted interactions. Since the electronic states
of the cluster Hamiltonian have no dispersion, it is sufficient to calculate them at a single point in the Brillouin zone (BZ), ${\bf k}_{_{\Gamma}}=(0,0,0).$

To link the current discussion with the available slab calculations, we
perform as a final step a set of calculations employing a $p(2\times 2)$ 
slab with five Si layers, \Fref{fig:1}~(b), similar
to~\Rref{Pehlke95}. The $k$-space integration is performed using 16 ${\bf
  k}_{\|}$ points in the whole surface Brillouin zone of the $2\times2$ unit cell.

Both cluster and slab calculations are carried out using the local-density
approximation (LDA) to the exchange-correlation functional, as obtained form
the Monte Carlo results by Ceperley and Alder~\cite{Ceperley80} in the
parameterization of Perdew and Zunger~\cite{Perdew81}.
For comparison all calculations are repeated with gradient corrections
as described by the BP, PW91 and BLYP functionals. 
Our choice to include these functionals in the test is partially motivated
by the good agreement between previous slab calculations and the experiment,
but on the other hand by the shortcomings of the BP and PW91 functionals,
cf.~\Rref{Nachtigall96}, found in the case of small systems.
DFT {\it plane-wave} calculations employing either slabs or Si$_9$ or larger
clusters and the BLYP functional have not yet been reported for the
\hm/Si(001) reaction energetics.

We employ norm-conserving $sp$-nonlocal
pseudopotentials for Si atoms, generated from a separate all-electron 
calculation of the Si atom for each exchange-correlation
functional~\cite{Fuchs} according to Hamann's scheme~\cite{Hamann89}. 
For gradient-corrected DFT calculations, the use of these consistently
constructed pseudopotentials ensures the proper description 
of core-valence exchange within each of the gradient-corrected functionals 
used~\cite{Fuchs&Pehlke}.
For the hydrogens passivating Si dangling bonds, a $s$-nonlocal pseudopotential
is generated following the Troullier and Martins
prescription~\cite{Troullier91}. However, for the hydrogen atoms taking part
in the reaction, we employ the full $1/r$ potential. 
For all energies quoted in the paper, plane waves with a kinetic energy up to 
$E_{\rm cut}= 30$~Ry (408 eV) were included in the basis set. 
While geometries and relative energies for systems consisting entirely of Si 
atoms are well converged already at 18 Ry (245 eV), the high quality basis set
is required for a correct description of the hydrogen wave functions close to the core. 

\subsection{Structure optimization}

We perform separate structure optimizations for each cluster size and for 
the LDA, BP, PW91 and BLYP functionals. The first step in all structure
relaxation runs was to optimize the positions of the terminating hydrogens,
with the subsurface silicon coordinates kept fixed at their bulk values, and
the symmetric dimer bond length set to $d=2.28$~\AA\ as found in~\Rref{Kratzer95}.

One of our present goals is to investigate to what extent clusters are
appropriate for a description of the Si(001) surface.  
One important feature of the clean surface is the buckling of the Si surface 
dimers (see the next Section). 
It is therefore interesting to  test if the buckling can be
modeled by clusters of appropriate size.
The failure of small clusters to reproduce the buckled surface structure is
sometimes attributed to the neglect or underestimate of elastic interactions 
between Si dimers in this approach. 
However, the importance of elastic interactions is established only for the 
alternation of buckling angle ('anti-ferromagnetic ordering') in the 
$p(2 \times 2)$ or $c(4 \times 2)$ reconstruction~\cite{Pehlke94}.
To distinguish elastic from electronic effects, we employ 
a two-step procedure  to determine the equilibrium structures of the 
bare clusters. First we sample the total energy as a function of the dimer
buckling angle $\alpha$, as done, for example, in
Refs.~\onlinecite{Kratzer95,Dabrowski92}, but without relaxation of any of the
deeper layers. Thus any elastic interactions between Si dimers are
avoided. Neighboring dimers (if there are any) are buckled in opposite directions, with buckling angles $\alpha$ and $-\alpha$, to mimic the $p(2\times 2)$ surface reconstruction.

The structure of the bare clusters is finally determined by
unconstrained relaxation of the topmost two Si layers.
Here we use the energy minimum as a function of $\alpha$ determined in 
the previous calculations as input for the starting geometry. 
The two pairs of hydrogens
saturating the second layer Si bonds in the $[1{\overline 1}0]$ direction, 
which would correspond to adjacent Si dimers on the Si(001) surface, are 
also allowed to relax. We have also tested full relaxation of all layers in
the case of the \si{9}{12} and \si{15}{16} clusters within LDA. 
The lack of any geometric constraints, however, tends to overestimate the
surface relaxation and introduces unrealistic atomic displacements. Full
relaxation would only be appropriate for clusters studied as objects in their
own interest, rather than as an approximation to the Si(001) surface.

The structure of the monohydride phase is determined by relaxing the adsorbate
and the two Si layers beneath it. In the case of the Si$_{21}$ cluster the
\hm\ molecule is adsorbed on the middle dimer.  
The geometries of transition states (TS) are determined by a search 
algorithm  using the ridge method proposed by Ionova and
Carter~\cite{Ionova93}. Since geometries are less sensitive to the qualtity of
the basis set than the total energies, we have found it sufficient to perform
geometry optimizations at a plane-wave cut-off of $E_{\rm cut}=18$~Ry. 
The structures are considered converged when all forces are
smaller than 0.05~eV/\AA.


\section{The clean \siloo surface}
\label{sec:III}

On the Si(001) surface, the surface Si atoms form dimers, leading to the 
(symmetric) $2\times 1$ reconstruction. 
The surface can reduce its symmetry if the Si surface atoms relax to 
different heights, {\it i.e.} the dimers are buckled. 
This leads to the formation of lower symmetry patterns, the 
asymmetric $p(2\times 1),$ $p(2\times 2),$ $c(4\times2)$. 
These reconstructions are characterized by the dimer buckling
angle $\alpha,$ the dimer bond length $d,$ and the energy favor per dimer
$\Delta E$ with respect to the symmetric $2\times 1$ reconstruction.

The calculations for \hm\ desorption from Si(001) using slabs together 
with DFT and those based on many-body wave functions (like CI) already 
differ in their description of the clean surface geometry. In the latter
calculations, mostly the \si{9}{12} cluster with one symmetric Si dimer was
used as reference state for the clean Si(001) surface. Radeke and Carter
verified that the symmetric cluster is the geometric ground state at the level
of theory used in their work~\cite{Radeke96}. Recent theoretical
studies~\cite{Konecny97,Yang97,Paulus} on Si clusters come to conflicting conclusions and thus leave the discussion about the ground state symmetry of Si clusters still open.

On the other hand, the DFT slab
calculations~\cite{Pehlke95,Kratzer95,Dabrowski92,Roberts90,Kruger} are
consistent in their predictions and show good agreement with recent
experiments. Direct evidence for the buckling comes from low-temperature STM
images~\cite{Wolkow92} and structure determinations by LEED~\cite{Over}.
Furthermore, observed core-level shifts \cite{Landemark} are inconsistent with 
symmetric surface dimers, but can be explained by buckled
dimers~\cite{Pehlke_SCLS}. The buckled surface reconstruction has gained further
support by the good agreement between the measured dispersion of surface
band states~\cite{Johansson} with calculations using the GW approximation~\cite{Northrup93}. 

The different symmetries of reconstruction correspond to different 
electronic structures at the surface. Previous theoretical
work~\cite{Pehlke95,Kruger,Liu95} has established the following picture:   
After dimerization of the surface Si atoms, they would both remain to have 
dangling bonds occupied by one electron, {\it i.e.} a degenerate 
electronic ground state. There are two principal possibilities for a lowering
of the electronic energy. Firstly, a bonding and antibonding linear
combination of orbitals could be formed (similar to a $\pi$-bond in a free
dimer), only one of which is occupied. In this case the Si surface dimer would
remain symmetric. The second possibility is a Jahn-Teller-like splitting of
the degeneracy. By buckling the Si dimer, the lower Si atom comes to an almost
planar bonding configuration with its three neighbors, while the upper Si atom
reaches a pyramidal configuration. A rehybridization of the orbitals at each
of the Si atoms results in a lowering of the dangling orbital at the upper Si
atom, and in an up-shift of the orbital at the lower Si atom, which is
accompanied by a transfer of electron density from the lower to the upper atom.The preferred way of stabilization depends on several factors: the possible 
strength of the $\pi$-bond, the possible energy gain due to rehybridization, 
the ability of the system to screen the increased Coulomb repulsion in 
the dangling bond of the upper Si atom, and the energetic cost of elastic
deformation in the deeper layers induced by the buckling. The ground state
geometry is thus determined by an interplay of both elastic and electronic
effects and could be quite sensitive to  different surface reconstructions 
and computational methods. 

Before we describe our own results for the ground state of clusters, we briefly
discuss the recent literature. Yang~\etal\cite{Yang97} do not find buckling
for the \si{9}{12} cluster, either within Hartree-Fock or in a DFT
calculation employing the B3LYP functional. Increasing the cluster size to
\si{15}{16}, however, unambiguously shows an energy favor for the 
buckled $p(2\times 2)$ reconstruction, $\sim$~3--5 kcal/mol (0.15--0.23~eV) per dimer depending on the basis set used. Kone\v{c}n\'y and
Doren~\cite{Konecny97} have used the same cluster and the BLYP functional to
study H$_2$O adsorption on \siloo and found a buckled dimer structure for \si{9}{12} with $\alpha=9.6^{\circ},$
$\Delta E= 0.05\mbox{ kcal/mol}$ (0.002~eV) and $d=2.27$ \AA\, while for the
two-dimer cluster \si{15}{16} they obtained $\alpha=15^{\circ},$ $\Delta E=
1.5\mbox{ kcal/mol}$ (0.07~eV) per dimer and $d=2.33$~\AA. The two research
groups use different relaxation constraints and basis sets. This could explain
the differences, given that a delicate balance of several effects is
responsible for the ground state configuration of the \si{9}{12} cluster.
%
%
\begin{center}
\begin{figure}[h]
\hskip0.75cm\psfig{figure=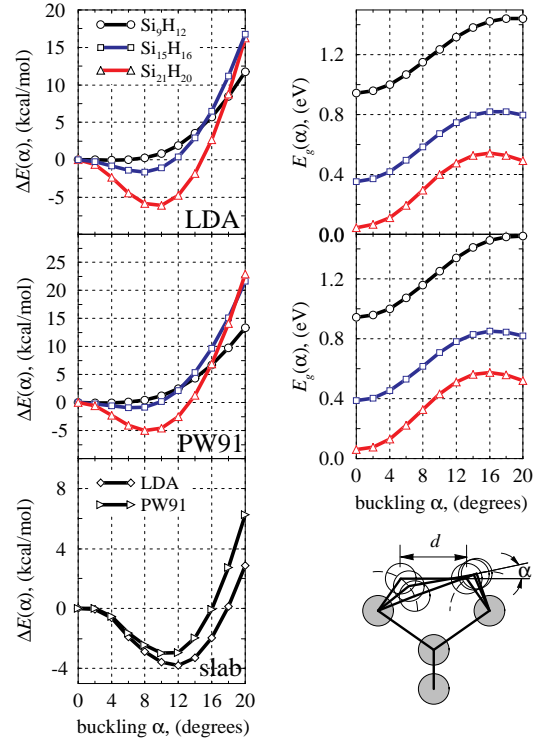,width=7cm}
\vskip0.5cm
\caption{ Total energy  preference $\Delta E = E(\alpha)-E_0$ (left column)
  and the gap $E_g(\alpha)$ between HOMO and LUMO (right column) for the
  clusters as a function of dimer buckling
  angle $\alpha.$ The bottom left panel shows $\Delta E(\alpha)$ for the
  five-layer $p(2\times 2)$ slab. $E_0$ is taken to be the energy of the
  corresponding unrelaxed symmetric ($\alpha = 0$) cluster/slab at given
  exchange-correlation functional. Buckling is achieved by gliding the
  two dimer Si atoms along arcs (dashed lines) as shown in the inset (bottom-right panel) and keeping all other atoms fixed (shaded circles).
\label{fig:2}}
\end{figure}
\end{center}

In the present study, the ground state of the clusters is determined in a 
two-step procedure. In the first step, the Si dimers are tilted as a whole,
while keeping the other cluster atoms fixed and preserving the length of the
dimer backbonds. We sample the total energy as a function of the dimer
buckling angle $\alpha$. The results for LDA and the PW91 functional are shown
in~\Fref{fig:2}. They are summarized in the 'pre-relaxation' column
of~\Tref{tab:1} for all functionals used in the calculations. 

The $\Delta E$ {\it vs.} $\alpha$ curves are
mainly affected by the cluster size, rather than by 
the approximation used for exchange and correlation. 
Upon increasing the cluster size, the minima become well pronounced 
and are shifted to larger $\alpha$ values, the upper bound being set by
$\alpha^{^{\rm (LDA)}}.$ In the case of \si{9}{12}, $\alpha=0$ is the only
minimum when using the BP and BLYP functionals. The values obtained within LDA
and PW91 for this cluster are small. Thus the \si{9}{12} cluster gives no
conclusive answer to the question whether the Si surface dimers are buckled or
not. The larger clusters, however, clearly show a preference for buckling.
Since we have frozen the elastic degrees of freedom, 
the only driving force that can lead to dimer buckling is rehybridization and
charge transfer. The pre-relaxation study supports the view that 
surface dimer buckling is mainly driven by electronic effects, 
while elastic effects are responsible for the alternation of  
the buckling angle.
%
%
\begin{table}
\caption{Parameters of the clusters and the slab for different
  exchange-correlation functionals; the energy favor for buckling $\Delta E$
  is given in kcal/mol per dimer and the gap $E_g$ between HOMO and LUMO  in
  eV. $d$ refers to the Si-Si dimer bond length in \AA\ and $\alpha$---to its
  buckling angle in degrees. 
\label{tab:1}}
\begin{minipage}{\columnwidth}
 \begin{tabular}{rrccrccc}
& \multicolumn{3}{c}{\rm pre-relaxation} & \multicolumn{4}{c}{\rm two layers
   relaxed$^a$\footnote[0]{\vskip-1.5cm $^a$ $\alpha$ and $d$ for the
     \si{21}{20} cluster correspond respectively to the buckling angle and bond length of the middle dimer.} 
   }\\ \cline{2-4}\cline{5-8}
& $\alpha$ & $|\Delta E|$ & $E_g$ & $\alpha$ & $|\Delta E|$ & $E_g$ & $d$
 \\ \hline 
LDA         &    &     &      &      &     &      &       \\ 
Si$_9$      & 4  & 0.1 & 1.00 & 6.9  & 0.1 & 1.18 & 2.21  \\  
Si$_{15}$   & 8  & 0.8 & 0.59 & 15.7 & 2.7 & 0.86 & 2.28  \\ 
Si$_{21}$   & 10 & 2.0 & 0.40 & 18.6 & 4.0 & 0.70 & 2.35  \\
slab        & 12 & 1.9 &      & 18.9 & 4.5 &      & 2.36  \\
\hline
BP           &    &     &      &      &     &      &      \\ 
Si$_9$       & 0  & --- & 0.99 & 0    & --- & 0.96 & 2.27 \\  
Si$_{15}$    & 6  & 0.4 & 0.54 & 15.3 & 2.6 & 0.86 & 2.37 \\ 
Si$_{21}$    & 8  & 1.4 & 0.33 & 18.1 & 4.4 & 0.78 & 2.47 \\
slab         & 10 & 1.3 &      & 18.6 & 5.1 &      & 2.46 \\
\hline 
PW91         &    &     &      &      &     &      &      \\
Si$_9$       & 2  & 0.0 & 0.96 & 4.9  & 0.1 & 1.10 & 2.24 \\ 
Si$_{15}$    & 6  & 0.5 & 0.43 & 15.7 & 3.1 & 0.88 & 2.32 \\ 
Si$_{21}$    & 8  & 1.7 & 0.33 & 18.0 & 4.6 & 0.74 & 2.40 \\ 
slab         & 10 & 1.5 &      & 18.3 & 5.1 &      & 2.40 \\
\hline
BLYP         &    &     &      &      &     &      &      \\
Si$_9$       & 0  & --- & 1.02 &  0   & --- & 0.99 & 2.27 \\ 
Si$_{15}$    & 4  & 0.1 & 0.51 & 12.2 & 1.6 & 0.72 & 2.33 \\ 
Si$_{21}$    & 6  & 0.8 & 0.27 & 16.6 & 3.5 & 0.72 & 2.44 \\ 
slab         & 10 & 0.8 &      & 16.6 & 4.7 &      & 2.43 \\
\end{tabular}
\end{minipage}
\end{table}

\begin{center}
\begin{figure}[h]
\hskip0.3cm\psfig{figure=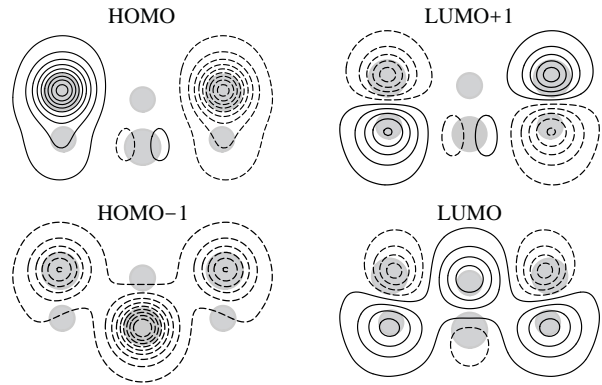,width=8.0cm}
\vskip0.5cm
\caption{ Top view of the \si{21}{20} cluster frontier orbitals.
The contour plots are taken in the (001) plane 0.9~\AA\ above the
buckled-up Si atom of the middle dimer (dimers are denoted by shaded circles). The wave functions are real-valued. Full (dashed) contour lines indicate positive (negative) sign. 
\label{fig:3}}
\end{figure}
\end{center}

The electronic structure of the cluster is influenced substantially by the 
dimer buckling. As a measure of the differences, we report the splitting
between the highest occupied and lowest unoccupied molecular orbitals (HOMO
and LUMO) of the cluster, $E_g$, in the right column in~\Fref{fig:2}. 
For \si{9}{12} $E_g$ is a monotonically increasing function of $\alpha$. 
It reaches its maximum for the largest used value of $\alpha=20^{\circ}$, 
where the two dimer silicons and the two second layer neighbors of the
buckled-down Si atom are almost coplanar. 
We attribute the opening of the HOMO-LUMO gap to changes in the symmetry
character of the orbitals.
While at $\alpha=0$ the HOMO and LUMO orbital correspond to the $\pi$ and 
$\pi^{*}$ orbital of the Si dimer, they gradually develop into orbitals
localized at one side of the Si dimer for increasing $\alpha.$ 
The orbital at the lower Si atom acquires more $p$-character and is lifted in
energy, while the orbital at the higher Si atom gets more $s$-character and is
energetically lowered~\cite{Liu95}. 
With increasing cluster size $E_g$ is strongly reduced, while the dependence on
$\alpha$ is almost unchanged. 
For the larger clusters we observe that the HOMO (LUMO) wave functions are no
longer localized at a single Si atom, but are linear combinations of the
dangling orbitals of all the buckled-up (buckled-down) Si atoms of the 
cluster. 
In the \si{21}{20} cluster, this leads to a splitting of the HOMO and LUMO 
levels into symmetric and antisymmetric states with respect to the mirror plane
in the cluster ~\Fref{fig:3}. Thus this cluster reflects already to some extent
the dispersion of surface bands observed in the slab.    
We expect that these changes in the electronic structure will also be reflected
in the chemical reactivity, {\it i.e.} in the adsorption barrier $E_{a}^{\rm
ads}$ for \hm\ molecules. It appears that a representation of the surface band
structure on the cluster level is required for a correct description of the
reaction energetics. 
%
%

Relaxing the first and second silicon layers does not qualitatively change the
results from the pre-relaxation study. However, the ground state energy and
geometry differ substantially. The buckled dimer configurations of the 
\si{15}{16} and the \si{21}{20} clusters are now more stable by 
1.5--3 kcal/mol (0.07--0.13~eV) and 3.5--4.5 kcal/mol (0.15--0.20~eV) per dimer, respectively.
The HOMO-LUMO gap is found to decrease with increasing cluster size.
The reduction can be partly attributed to a splitting of both the HOMO
and LUMO state  due to linear combinations of dangling
bonds at neighboring Si dimers in the larger clusters (see Fig. 3).
However, a decrease is also observed, in particular at $\alpha=0$,
for a suitably averaged gap $\left<E_g \right>$, defined as 
$$
\left< E_g(\alpha) \right>_{_N} =
\frac{1}{N}\sum\limits_{j=0}^{N-1}[E_{_{{\rm LUMO}+j}}(\alpha)-E_{_{{\rm HOMO}-j}}(\alpha)],
$$
where $N$ is the number of dimers in the cluster.
The reduction of the gap must be attributed to a weakening of the
$\pi$-bonding with increasing cluster size.
As could be anticipated, \si{21}{20} gives
closest agreement with the five-layer $p(2\times 2)$ slab. The \si{9}{12}
cluster, on the contrary, is not large enough to model the properties
of the clean \siloo surface. Buckling in this case is strongly influenced by
relaxation constraints and the approximation to exchange and correlation. 
While for the two larger clusters all functionals used here give similar
results, their predictions for the \si{9}{12} ground state symmetry are
qualitatively different. Though LDA and PW91 favor buckling with
$\alpha\approx7^{\circ}$ and $5^{\circ}$, respectively, $|\Delta E|$ is too
small ($\sim 0.1$~kcal/mol) to allow us to conclude unequivocally about the
\si{9}{12} cluster symmetry. 

For the two-dimer \si{15}{16} cluster using the BLYP functional we get $d$ and $\Delta E$
values identical to those of Kone\v{c}n\'y and Doren, but their
predicted \si{9}{12} geometry is at variance with ours. Possibly the
overestimation of the surface relaxation due to the absence of
any geometric constraints in \Rref{Konecny97} is more crucial for the
\si{9}{12} cluster than for the larger ones. 

At our imposed relaxation constraints the BP and
BLYP functionals are found to give results for \si{9}{12} in agreement with
the CI methods~\cite{Radeke96,Paulus}. For the \si{15}{16} cluster, however,
buckling is always energetically favorable within DFT, whereas recent
multi-reference CI calculations~\cite{Paulus} find the symmetric cluster
to be lowest in energy. At present it is not clear if this is due to a lack of
the CI calculations to recover the full correlation energy, or due to an 
inadequacy of the exchange-correlation functionals we are using.
For the largest \si{21}{20} cluster, we are not aware of CI calculations
addressing the dimer buckling.

\section{Reaction energetics}
\label{sec:IV}

The energetics of dissociative adsorption and associative desorption 
of \hm\ is characterized by three
points along the reaction pathway, the energies of the structures corresponding to the
monohydride phase $E_{11},$ the transition state energy $E_{\rm TS}$ and 
the sum of bare cluster energy $E_{00}$ and that of the free \hm\ molecule. 
Hence, the
quantities of interest are defined by the differences 
\begin{eqnarray}
E_{a}^{\rm ads} & = & E_{\rm TS}-(E_{00}+E({\rm H}_2)),\nonumber \\
E_{a}^{\rm des} & = & E_{\rm TS}-E_{11},               \nonumber \\
E_{\rm rxn}     & = & E_{00}+E({\rm H}_2)-E_{11}.      \nonumber
\end{eqnarray} 
%
%
\begin{center}
\begin{figure}[h]
\hskip1cm\psfig{figure=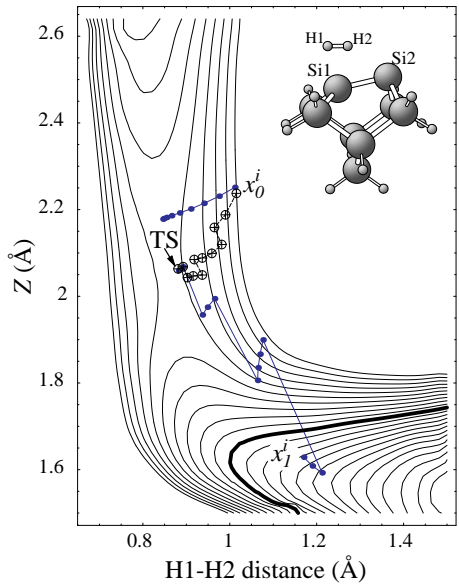,width=6cm}
\vskip0.5cm
\caption{ Sketch of the ridge method search for TS and the adiabatic PES for
  the Si$_9$ cluster within LDA. Crossed circles ($\oplus$) denote the
  successive approximations to the TS. Contour spacing is 0.05 eV and energy
  is measured with respect to $E_{00}+E({\rm H}_2)$ (solid contour
  line). Transition state  geometry is shown in the inset. PES is plotted for
  \hm\ molecule impinging  perpendicularly to the surface with its center of
  mass right above the Si1 site, $Z$ being the distance between them.
\label{fig:4}}
\end{figure}
\end{center}

The monohydride geometry of each cluster is obtained by saturating 
the dangling bonds of one Si dimer with H atoms. 
The equilibrium Si-H bond is 1.52~\AA\ within LDA, PW91 and BLYP. When using
the BP functional a bond length larger by 0.03~\AA\ is obtained. 
The changes in $d$ of the hydrogenated dimer are mainly governed by the
approximation to the exchange and correlation, while the cluster size accounts
only for differences of $\sim 0.01$~\AA\ for a  given functional.
We note that our TS and monohydride structures imply reaction of a single \hm\ 
molecule with the Si(001) surface. For comparison to experimental data taken
at finite hydrogen coverage, information about the coverage dependence of the energetics is also required.  
A single \hm\ molecule per $p(2\times2)$ unit cell in the
extended slab case corresponds to a coverage of $\Theta= 0.5$ monolayers
(ML). To assess the coverage dependence of the reaction energy 
$E_{\rm rxn}$, test calculations were carried out for
 a completely covered slab, $\Theta=1$~ML. 
To obtain a better understanding of the role of buckling for the
reaction energetics, we also performed test calculations for the 
symmetric $p(2\times1)$ reconstructed Si(001) surface as possible 
reference of the clean surface.
We note that the occupation of adjacent Si dimers by hydrogens 
could have a (probably small) influence on the transition state 
geometries and energies via interaction with the neighboring monohydride. 
This effect can be studied to some extent with the help of the 
two- and three-dimer clusters, which allow for adjacent doubly occupied Si
dimers. To get some insight into the influence of neighboring monohydrides on
the asymmetric TS, we have performed calculations with \si{15}{16+x} and
\si{21}{20+x} clusters with $x=4$ and 6, respectively. 
%
%
\begin{table}[t]
\caption{Transition state geometry parameters in \AA\ (see the inset
  in~\protect{\Fref{fig:4}}). $\alpha'$ denotes the buckling angle in degrees of the unoccupied dimer(s) in the case of slab and Si$_{15},$ Si$_{21}$ clusters.
\label{tab:2}}
\begin{tabular}{rccccccc}
& $R_{_{\rm H1-H2}}$ &$R_{_{\rm Si1-H1}}$ & $R_{_{\rm Si1-H2}}$
&$R_{_{\rm Si2-H2}}$ &$d$ & $\alpha_{_{\rm TS}}$ & $\alpha'$\\ \hline 
\multicolumn{8}{l}{LDA} \\
           Si$_9$     & 0.88 & 2.12 & 2.10 & 2.38 & 2.30 & 13.7 & ---  \\  
           Si$_{15}$  & 0.97 & 1.93 & 1.98 & 2.24 & 2.38 & 12.1 & 15.6 \\
           Si$_{21}$  & 1.06 & 1.95 & 2.14 & 2.09 & 2.38 & 13.0 & 15.5 \\
           slab       & 0.93 & 2.00 & 2.11 & 2.24 & 2.39 & 13.5 & 19.0 \\
\hline
\multicolumn{8}{l}{BP}\\
          Si$_9$      & 0.94 & 1.96 & 1.95 & 2.34 & 2.42 & 12.2 & ---  \\  
          Si$_{15}$   & 1.02 & 1.84 & 1.95 & 2.19 & 2.46 & 11.5 & 15.5 \\ 
          Si$_{21}$   & 1.02 & 1.82 & 2.01 & 2.12 & 2.47 & 11.6 & 14.8 \\
          slab        & 1.03 & 1.80 & 1.94 & 2.11 & 2.50 & 13.1 & 18.7 \\
\hline
\multicolumn{8}{l}{PW91}\\
            Si$_9$    & 0.89 & 1.98 & 1.96 & 2.32 & 2.37 & 12.2 & ---  \\ 
            Si$_{15}$ & 0.99 & 1.84 & 1.94 & 2.19 & 2.40 & 10.7 & 14.6 \\ 
            Si$_{21}$ & 1.08 & 1.89 & 2.11 & 2.04 & 2.41 & 13.6 & 14.9 \\ 
            slab      & 0.99 & 1.83 & 1.99 & 2.12 & 2.43 & 12.7 & 18.4 \\
\hline
\multicolumn{8}{l}{BLYP}\\
            Si$_9$    & 0.91 & 1.93 & 1.95 & 2.31 & 2.40 & 11.9 & ---  \\ 
            Si$_{15}$ & 0.96 & 1.86 & 1.98 & 2.23 & 2.44 & 10.2 & 12.6 \\ 
            Si$_{21}$ & 0.96 & 1.83 & 2.02 & 2.11 & 2.48 & 12.1 & 13.8 \\ 
            slab      & 1.02 & 1.74 & 1.89 & 2.12 & 2.48 & 11.1 & 18.4 \\
\end{tabular}
\end{table}

In this study, we concentrate on the pre-pairing scenario for the \hm\ 
reaction with the Si(001) surface. Consequently, we locate the asymmetric TS 
of \hm\ desorption from a single Si dimer for all clusters and functionals
used in this study. In principle this can be achieved by mapping out the
related potential-energy surface (PES), like e.g. in \Rref{Pehlke95}.
As an example, the potential energy as a function of the distance between the
two H atoms and the H$_2$-cluster distance $Z$ is shown in~\Fref{fig:4}. For
each configuration of the two H atoms, the Si atoms in the two topmost layers
have been relaxed. They follow 'adiabatically' the motion of the H atoms. Thus
we make sure that the lowest possible TS in the multidimensional space of all
mobile atomic coordinates is found.
An alternative and generally faster approach uses a search algorithm, thus
avoiding the need to map out all the points in the PES. All degrees of freedom
of the \hm\ molecule and the topmost two cluster/slab Si layers are included in
the search. 
The ridge method~\cite{Ionova93} implemented here 
starts from a pair of coordinates 
$x_0$ and $x_1$  in the multi-dimensional configuration space of the system, 
which denote the reactants (\hm\ above the bare surface) and the products (the
monohydride), respectively. An iterative search is then performed by halving
the interval $[x_{0}^{i}, x_{1}^{i}]$ in each step. To reduce the number of
required steps, we have shifted the input coordinates towards some initial guess $(x_{0}^{i},x_{1}^{i})$ closer to the TS. 
A projection of the search path onto a two-dimensional slice of the 
coordinate space is shown in~\Fref{fig:4} for the \si{9}{12+2} cluster within
LDA. As seen, there is excellent agreement between the two schemes. 

The geometries of the calculated TS are collected
in~\Tref{tab:2}. As expected, the H-H bond is stretched in the asymmetric TS configuration of all \si{x}{y+2} clusters, with $R_{_{\rm H1-H2}}$ being
largest for dissociation on the Si$_{21}$ cluster. 
As a consequence, the atom H2 at the
transition state is by 0.2--0.3~\AA\ closer to the buckled-up Si2 atom as
compared with the Si$_9$ cluster. The presence of two H atoms over
the Si1 site partially blocks the mechanism that leads to anticorrelated
dimer buckling and therefore $\alpha_{_{\rm TS}} < \alpha$. For the 
Si$_{21}$ cluster the unoccupied dimers are somewhat affected by the
adsorption event, but their buckling angle $\alpha'$ is only a few degrees
smaller than that of the clean surface. 
This small change is due to the dimer row termination by hydrogens in the 
clusters, and is absent in the slab geometries, where $\alpha'\approx\alpha.$
%
%
\begin{center}
\begin{figure}[h]
\hskip1.5cm\psfig{figure=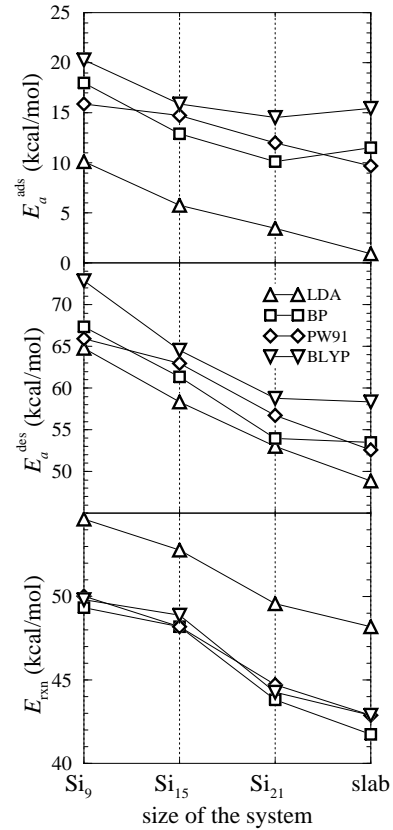,width=5cm}
\vskip0.5cm
\caption{\hm/Si(001) reaction energetics as a function of the size of the
  system used to model the \siloo surface (see also~\protect{\Tref{tab:3}}).
\label{fig:5}}
\end{figure}
\end{center}

Most noticeable in the results shown in~\Fref{fig:5} (numerical values are
compiled in~\Tref{tab:3}) is a clear dependence of
$E_{a}^{\rm ads},$ $E_{a}^{\rm des}$ and $E_{\rm rxn}$ on cluster size.
The common trend for all functionals is an effective flattening of PES 
along the reaction path with increasing cluster size.
Both the reaction energy and the adsorption barrier are reduced. 
In general, for all clusters considered here, LDA gives a lower bound
for the activation barriers and an upper one for $E_{\rm rxn}.$
For a given size of the cluster, we can compare the effect of the
exchange-correlation functional employed in the calculation.
As far as $E_{\rm rxn}$ is concerned, all gradient-corrected functionals behave
in a very similar way (see bottom panel of~\Fref{fig:5}). This gives
credibility to the statement that gradient-corrected DFT yields an accurate description of reaction energies.
The quantities $E_{a}^{\rm ads},$ $E_{a}^{\rm des}$, 
involving transition state energies, show a stronger variation.  
The differences between BP and PW91 are only $\sim 2$~kcal/mol, with the sign
depending on the size of the system. 
The BLYP functional gives the highest value for $E_{a}^{\rm ads, des}$ 
of all tested functionals. Since similar performance has been already
established for small systems~\cite{Nachtigall96} and the Si$_9$
cluster~\cite{Pai95}, our calculations confirm these results for extended clusters and slabs.
%
%
\begin{table}
\caption{Activation and reaction energies in kcal/mol (eV) per molecule for \hm\  adsorption/desorption on \siloo surface via pre-pairing mechanism (ZPE
  not included). 
\label{tab:3}}
\begin{tabular}{rccc}
 & $E_{a}^{\rm ads}$ & $E_{a}^{\rm des}$ & $E_{\rm rxn}$\\ \hline 
\multicolumn{4}{l}{LDA}\\
 Si$_9$    & 10.2 (0.44) & 64.9 (2.81)  & 54.7 (2.37)  \\
 Si$_{15}$ &  5.7 (0.25) & 58.4 (2.53)  & 52.7 (2.29)  \\
 Si$_{21}$ &  3.5 (0.15) & 53.1 (2.30)  & 49.6 (2.15)  \\ 
 slab      &  0.9 (0.04) & 49.0 (2.12)  & 48.1 (2.09)  \\ \hline
\multicolumn{4}{l}{BP}\\ 
 Si$_9$    & 18.1 (0.78) & 67.3 (2.92)  & 49.3 (2.14)   \\
 Si$_{15}$ & 13.3 (0.56) & 61.4 (2.66)  & 48.1 (2.09)   \\
 Si$_{21}$ & 10.2 (0.44) & 53.9 (2.34)  & 43.8 (1.90)   \\ 
 slab      & 11.6 (0.50) & 53.4 (2.32)  & 41.8 (1.81)   \\ \hline
\multicolumn{4}{l}{PW91}\\
 Si$_9$    & 15.9 (0.69) & 66.0 (2.86)  & 50.1 (2.17)   \\
 Si$_{15}$ & 14.8 (0.64) & 63.0 (2.73)  & 48.2 (2.09)   \\
 Si$_{21}$ & 12.0 (0.52) & 56.7 (2.46)  & 44.7 (1.94)   \\ 
 slab      &  9.6 (0.42) & 52.5 (2.28)  & 42.9 (1.86)   \\ \hline
\multicolumn{4}{l}{BLYP}\\
 Si$_9$    & 20.2 (0.88) & 72.8 (3.16)  & 49.7 (2.16)   \\
 Si$_{15}$ & 15.8 (0.69) & 64.6 (2.80)  & 48.8 (2.12)   \\
 Si$_{21}$ & 14.5 (0.63) & 58.8 (2.55)  & 44.3 (1.92)   \\
 slab      & 15.5 (0.67) & 58.4 (2.53)  & 42.9 (1.86)   \\
\end{tabular}
\end{table}

The Si$_{21}$ cluster displays both the lowest adsorption
and desorption barriers. Comparison between this cluster and the slab 
shows that their predictions agree to within $\sim 3$~kcal/mol for all 
quantities. We conclude that the Si$_{21}$ cluster gives a fair
description of the Si(001) surface, while the others are inadequate approximations for
the surface. While the desorption barriers derived from the Si$_{21}$ cluster 
are in the range of $56\pm 3$~kcal/mol for all functionals, $E_{a}^{\rm ads}$ 
is more sensitive to the functional used, with values covering a range
of 11~kcal/mol. The adsorption barriers derived from the gradient-corrected
functionals using this cluster are 7--11~kcal/mol higher than the LDA
barrier. This is in accord with the established picture that LDA tends to
underestimate adsorption barriers at surfaces~\cite{Hammer}. 
We note that the Si$_{21}$ cluster yields higher barriers for the PW91 functional than for BP, while the slab calculations give the opposite result.  
%
%
\begin{center}
\begin{figure}[h]
\hskip0.3cm\psfig{figure=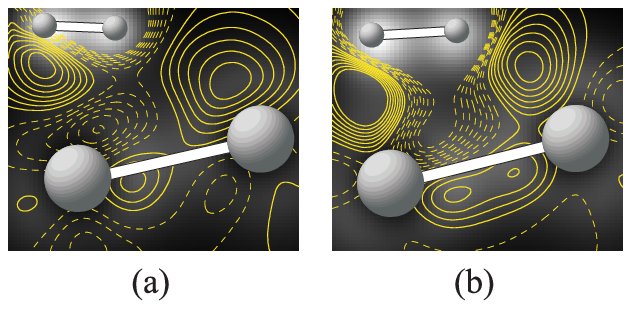,width=8cm}
\vskip0.5cm
\caption{Total valence electron density $n({\bf r})$ (grey shading) and the
 density difference $\Delta n({\bf r})$ (contour plot) in the plane
 containing \hm\ molecule and the Si-Si surface dimer at the TS
 geometries of the \si{9}{12+2} (a) and the \si{21}{20+2} cluster (b) calculated with the PW91 functional. Density difference is defined by $\Delta n({\bf r})= n({\bf r})-n_{_{\rm TS}}^{\rm clust}({\bf r})-n_{_{\rm TS}}^{\rm H_2}({\bf r}).$ The full contour lines correspond to $\Delta n>0$ and dashed lines to $\Delta n<0$. The plot levels are the same for both clusters.
\label{fig:6}}
\end{figure}
\end{center}

It is instructive to analyze why the smaller Si$_{9}$ and Si$_{15}$ clusters
give a poor description of the physics at the Si(001) surface, despite the
fact that the Si--H bound is localized and should therefore be well
represented already in the smallest cluster used. For $E_{\rm rxn}$, the
variation with cluster size is mainly due to differences in the bare clusters
which are used as reference states. The buckling of the Si dimers
characteristic for the Si(001) surface only fully develops in the largest
clusters, because it requires a correct description of the electronic surface
states. Thus non-local electronic effects enter the calculation of the
reaction energy. They give rise to a reduction of $E_{\rm rxn}$ by about 5~kcal/mol compared to the value obtained with the Si$_9$ cluster.
For $E_{a}^{\rm ads}$, the differences between small clusters and the slab
calculation are even larger. This is due to the fact that the dangling bonds
of the Si dimer act as frontier orbitals in the reaction with \hm. 
The importance of the change in the HOMO and LUMO position that accompany
dimer buckling has also been emphasized in recent first principles studies of
H$_2$O~\cite{Konecny97}, C$_2$H$_2$~\cite{Liu95}, BH$_3$~\cite{Konecny971} and 1,3-cyclohexadien~\cite{Konecny972} reactions with the Si(001) surface.
As our calculations show, the HOMO-LUMO gap $E_g$ is substantially reduced 
when going from the (mainly $\pi$-bonded) Si dimer in Si$_9$ to the 
surface band states in the slab. 
At the transition state for adsorption,
the \hm\ molecular orbitals, in particular the antibonding $^{3}\Sigma_{u}$
orbital, mix with both the HOMO and LUMO orbital, which is facilitated by a
small $E_g$. At the TS of the Si$_9$ cluster, the \hm\ mainly interacts with
one dangling bond of the lower Si atom (see Fig.~\ref{fig:6}). 
In the more extended systems with smaller $E_g$, there is also a direct
interaction between the \hm\ molecule and the upper Si atom. This shows up in
the more localized induced charge distribution between H2 and Si2 (right panel
of Fig.~\ref{fig:6}), and in the different geometrical structure of the transition state. 

The trends in the reaction energetics outlined above do not change
substantially if TS
and monohydride configurations of the Si$_{15},$ Si$_{21}$ clusters with more
than one monohydride are employed. Test calculations using
\si{15}{16+4} and \si{21}{20+6} clusters show that finite coverage effects at
the TS introduce a variation of less than 3~kcal/mol in the calculated
barriers. 
The coverage dependence of $E_{\rm rxn}$ was studied for the slab~\cite{Pehlke98}
with initial coverage $\Theta=1$~ML, \Tref{tab:4}. It is energetically more
expensive to desorb a \hm\ from one of the dimers if the other stays
monohydride. Thus, $E_{\rm rxn}(1\rightarrow 0.5\mbox{ ML})$ is about
3--4~kcal/mol larger than  $E_{\rm rxn}$ calculated for initial coverage
$\Theta=0.5$~ML. The reaction endothermicity for the removal of a whole
monolayer comes out as the average of the endothermicities associated with the
removal of each of the two monohydrides in the unit cell. 
It is interesting to note that $E_{\rm rxn}$ for the Si$_{21}$
  cluster with one monohydride (see~\Tref{tab:3}) coincides with 
$E_{\rm rxn}(1\mbox{ ML})$, rather than with the low-coverage limit of the
slab, $E_{\rm rxn}(0.5 \mbox{ ML})$.
The coverage dependence calculated for the slab shows the same trend as 
the experimental data by Flowers~\etal\cite{Flowers98}, where a slight
increase of the desorption energy with coverage is observed, $E_{a}^{\rm
  des}(\Theta) = (55.8+1.1\times\Theta)\mbox{ kcal/mol}$ for initial coverages
in the range 0.01--1~ML. 
%
%
\begin{table}
\caption{Coverage dependence of the reaction energy $E_{\rm rxn}(\Theta)$ for the slab.
\label{tab:4}}
\begin{tabular}{rccc}
& \multicolumn{3}{c}{$E_{\rm rxn}(\Theta),$ kcal/mol (eV) per \hm}  \\ \cline{2-4}
 & 0.5~ML & 1~ML & $1 \rightarrow 0.5$~ML\\ \hline 
LDA  & 48.1 (2.09) & 49.7 (2.16) & 51.3 (2.22)  \\ 
BP   & 41.8 (1.81) & 43.8 (1.90) & 45.7 (1.98)  \\ 
PW91 & 42.9 (1.86) & 44.7 (1.94) & 46.5 (2.02)  \\
BLYP & 42.9 (1.86) & 44.3 (1.92) & 45.7 (1.98)  \\
\end{tabular}
\end{table}

The reaction energy $E_{\rm rxn}$ calculated with respect to 
the symmetric $2\times1$ reconstructed Si(001) surface shows very similar
behavior. The respective $E_{\rm rxn}(\Theta)$ values are increased
by a few kcal/mol. In this case, when a \hm\ molecule desorbs, it leaves
behind a symmetric unoccupied dimer and therefore the final state is
$\sim\Delta E$ higher in energy than for the buckled surface reconstruction. 

\section{Conclusions}
\label{sec:V}

We have presented a systematic {\it ab initio} study of the \hm/Si(001) 
reaction energetics employing three
clusters in plane-wave DFT calculations. A five-layer
$p(2\times2)$ slab was used as a reference to analyze the convergence of
the cluster predictions. All calculations were performed within LDA and
with the non-local BP, PW91 and BLYP exchange-correlation functionals. As our
results show, a conservative conclusion can be drawn that the most frequently
used \si{9}{12} cluster is not large enough either to model the properties of
the bare Si(001) surface or the molecular \hm\ dissociation on it. The latter
stems from the fact that this cluster is not capable of recovering the
surface electronic structure. Though the \hm\ reaction with the Si(001) surface
is considered to be a highly localized event, non-local effects enter the
reaction energetics via their influence on the surface bands. Hence, one could
also expect different performance for the various functionals.

Our analysis shows that the quality of a given exchange-correlation
functional should not be assessed without referring to the particular size of
the cluster employed to study the \hm/Si(001) adsorption/desorption
process. Indeed, it is evident, by inspecting the BLYP section of~\Tref{tab:3}
for example, that one could infer for the Si$_9$ cluster an activation barrier
to desorption much higher than the experimental values. Hence, as usually
proceeded, the pre-pairing mechanism could be ruled out on energetic
grounds. Such a conclusion, however, seems to be premature if one refers to
the slab or, eventually, the Si$_{21}$ cluster prediction within the same functional.

In contrast to the one-dimer cluster approximation, the \si{21}{20} cluster
was found to be close in its predictions to the slab calculations in all
aspects of the reaction considered. With $E_{a}^{\rm des} = 56\pm 3$~kcal/mol
it is also well within the range of the experimentally determined desorption
barriers. The relatively large spread in the calculated energies comes from
the functionals used, with the BLYP functional giving the largest value for
$E_{a}^{\rm ads,des}$. ZPE corrections were not considered, but as they amount
to a few kcal/mol at most and are essentially the same for different
functionals, no qualitative change of our results is expected upon their
inclusion. Concluding, our findings suggest that the effect of the cluster
size in modeling the \hm\ reaction with the Si(001) surface is significant and
some of the previous works may well need a revision.

\acknowledgments

One of the authors (E.P.) is much indebted to A. P. Seitsonen and M. Fuchs for
the valuable discussions and help during this study.
We are grateful to Dr. E. Pehlke for critical reading of the manuscript and suggestions.



\begin{references}

\bibitem{Doren96} For a nice review see D. J. Doren, {\it Advances in Chemical
  Physics}, edited by I. Prigogine and S. A. Rice (Wiley, New York, 1996),
Vol. 95, p. 1; K. Kolasinski, \jour{Int. J. Mod. Phys. B}{9}{2753}{1995}.

\bibitem{Sinniah89} K. Sinniah, M. G. Sherman, L. B. Lewis, W. H. Weinberg,
J. T. Yates, Jr., and K. C. Janda, \jour{\PRL}{62}{567}{1989}.

\bibitem{Wise91} M. L. Wise, B. G. Koehler, P. Gupta, P. A. Coon, and
S. M. George, \jour{\SS}{258}{5482}{1991}.

\bibitem{Hofer92} U. H\"ofer, L. Li, and T. Heinz, \jour{\PRB}{45}{9485}{1992}.

\bibitem{Nachtigall93} P. Nachtigall, K. D. Jordan and C. Sosa, \jour{J. Phys. Chem.}{97}{11666}{1993}.

\bibitem{Nachtigall91} P. Nachtigall, K. D. Jordan and K. C. Janda, \jour{\JCP}{95}{8652}{1991}.

\bibitem{Wu93} C. J. Wu, I. V. Ionova, and E. A. Carter, \jour{\SS}{295}{64}{1993}.

\bibitem{Nachtigall94} P. Nachtigall, K. D. Jordan, and C. Sosa,
\jour{\JCP}{101}{8073}{1994}.

\bibitem{Jing93} Z. Jing and J. Whitten, \jour{\JCP}{98}{7466}{1993}.

\bibitem{Jing95} Z. Jing and J. Whitten, \jour{\JCP}{102}{3867}{1995}.

\bibitem{Nachtigall96} P. Nachtigall, K. D. Jordan, A. Smith and H. J\'onsson, \jour{\JCP}{104}{148}{1996}.

\bibitem{Radeke96} M. R. Radeke and E. A. Carter, \jour{\PRB}{54}{11803}{1996}; \jour{{\it ibid.}}{55}{4649}{1997}.

\bibitem{Pai95} S. Pai and D. Doren, \jour{\JCP}{103}{1232}{1995}.

\bibitem{Vittadini94} A. Vittadini and A. Selloni, \jour{Chem. Phys. Lett.}{235}{334}{1995}.

\bibitem{Pehlke95} E. Pehlke and M. Scheffler, \jour{\PRL}{74}{952}{1995}, mtrl-th/9412004.

\bibitem{Kratzer95} P. Kratzer, B. Hammer, and J. K. N{\o}rskov,
\jour{\PRB}{51 }{13 432}{1995}, mtrl-th/9503008.

\bibitem{Flowers93} M. C. Flowers, N. B. H. Jonathan, Y. Liu, and A. Morris, \jour{\JCP}{99}{7038}{1993}.

\bibitem{Bratu96} P. Bratu, W. Brenig, A. Gro$\beta,$ M. Hartmann, U. H\"ofer, P. Kratzer, and R. Russ, \jour{\PRB}{54}{5978}{1996}.

\bibitem{N&D} P. Nachtigall (unpublished); D. Doren (unpublished).

\bibitem{bX} A. D. Becke, \jour{Phys. Rev. A}{38}{3098}{1988}. 

\bibitem{lypC} C. Lee, W. Yang, and R. G. Parr, \jour{\PRB}{37}{786}{1988}.

\bibitem{pw91} J. P. Perdew, J. A. Chevary, S. H. Vosko, K. A. Jackson,
M. R. Pederson, D. J. Singh,  and C. Fiolhais, \jour{\PRB}{46}{6671}{1992}.

\bibitem{pC} J. P. Perdew, \jour{\PRB}{33}{8822}{1986}.

\bibitem{b3lyp} P. J. Stephens, F. J. Devlin, C. F. Chabalowski, and
M. J. Frisch, \jour{J. Phys. Chem.}{98}{11623}{1994}; A. D. Becke, \jour{\JCP}{98}{5648}{1993}.

\bibitem{Nachtigall95} P. Nachtigall and K. D. Jordan, \jour{\JCP}{102}{8249}{1995}.

\bibitem{Dabrowski92} J. D\c{a}browski and M. Scheffler, \jour{Appl. Surf. Sci.}{56--58}{15}{1992}

\bibitem{Wolkow92} R. A. Wolkow, \jour{\PRL}{68}{2636}{1992}.

\bibitem{Brenig} W. Brenig, A. Gross, and R. Russ, \jour{Z. Phys. B}{96}{231}{1994}.

\bibitem{Flowers98} M. C. Flowers, N. B. H. Jonathan, A. Morris,  and
 S. Wright, \jour{\JCP}{108}{3342}{1998}.

\bibitem{Bockstedte97} M. Bockstedte, A. Kley, J. Neugebauer, and M. Scheffler,
 \jour{Comp. Phys. Commun.}{107}{187}{1997};
\newblock see also http://www.fhi-berlin.mpg.de/th/fhimd/code.html .

\bibitem{Ceperley80} D. M. Ceperley and B. J. Alder, \jour{\PRL}{45}{567}{1980}.

\bibitem{Perdew81} J. P. Perdew and A. Zunger, \jour{\PRB}{23}{5048}{1981}.

\bibitem{Fuchs}M. Fuchs and M. Scheffler, Comput. Phys. Commun. (submitted).

\bibitem{Hamann89} D. R. Hamann, \jour{\PRB}{40}{2980}{1989}.

\bibitem{Fuchs&Pehlke} M. Fuchs, M. Bockstedte, E. Pehlke, and M. Scheffler,
\jour{\PRB}{57}{2134}{1998}.

\bibitem{Troullier91} N. Troullier and J. L. Martins, \jour{\PRB}{43}{1993}{1991}.

\bibitem{Pehlke94} J. Dabrowski, E. Pehlke, and M. Scheffler, 
\jour{\PRB}{49}{4790}{1994}.

\bibitem{Ionova93} I. V. Ionova and E. A. Carter, \jour{\JCP}{98}{6337}{1993}.

\bibitem{Konecny97} R. Kone\v{c}n\'y and D. J. Doren, \jour{\JCP}{106}{2426}{1997}.

\bibitem{Yang97} C. Yang, S. Y. Lee, and H. C. Kang, \jour{\JCP}{107}{3295}{1997}.

\bibitem{Paulus} B. Paulus, \jour{\SS}{408}{195}{1998}, cond-mat/9803219.

\bibitem{Roberts90} N. Roberts and R. J. Needs, \jour{\SS}{236}{112}{1990}.

\bibitem{Kruger} P. Kr\"uger and J. Pollmann, \jour{\PRL}{74}{1155}{1995}.

\bibitem{Over} H. Over, J. Wasserfall, W. Ranke, C. Ambiatello, R. Sawitzki, D. Wolf, and W. Moritz, \jour{\PRB}{55}{4731}{1997}. 

\bibitem{Landemark} E. Landemark, C. J. Karlsson, Y.-C. Chao, and 
R. I. G. Uhrberg, \jour{\PRL}{69}{1588}{1992}; E. L. Bullock, R. Gunnella, L. Patthey, T. Abukawa, S. Kono, C. R. Natoli, and L. S. O. Johansson, \jour{\PRL}{74}{2756}{1995}.

\bibitem{Pehlke_SCLS} E. Pehlke and M. Scheffler,
Phys. Rev. Lett. {\bf 71}, 2338 (1993).

\bibitem{Johansson} L. S. O. Johansson, R. I. G. Uhrberg, P. Martensson, and
G. V. Hansson, \jour{\PRB}{42}{1305}{1990}.

\bibitem{Northrup93} J. E. Northrup, \jour{\PRB}{47}{10032}{1993}.

\bibitem{Liu95} Q. Liu and R. Hoffmann, \jour{J. Am. Chem. Soc.}{117}{4082}{1995}.


\bibitem{Hammer} B. Hammer, K. W. Jacobsen, and J. K. N{\o}rskov,
\jour{\PRL}{70}{3971}{1993}; B. Hammer, M. Scheffler, K. W. Jacobsen, and
J. K. N{\o}rskov, \jour{{\it ibid.}}{73}{1400}{1994}. 

\bibitem{Konecny971} R. Konecny and D. J. Doren, \jour{J. Phys. Chem. B}{101}{10983}{1997}.

\bibitem{Konecny972} R. Konecny and D. J. Doren, \jour{J. Am. Chem. Soc.}{119}{11098}{1997}.
 
\bibitem{Pehlke98} For $E_{\rm rxn}(\Theta)$ in case of hydrogen chemisorption on vicinal Si(001) surfaces, a very similar coverage dependence was calculated,see E. Pehlke and P. Kratzer,  Phys. Rev. B, in press, cond-mat/9807329.

\end{references}
\end{document}